\documentclass[preprint,showpacs,preprintnumbers,amsmath,amssymb]{revtex4}

\usepackage{graphicx}
\usepackage{dcolumn}
\usepackage{bm}
\usepackage[]{ams,amsmath,amssymb,epsfig,color}
\newcommand{\be}{\begin{equation}}
\newcommand{\ee}{\end{equation}}
\begin{document}


\title{Chaos in generalized Jaynes-Cummings model. Kinetic approach}

%

%
\author{ L. Chotorlishvili, Z. Toklikishvili }

\affiliation{Physics Department of the Tbilisi State
University,Chavchavadze av.3,~~0128, Tbilisi, Georgia}

\affiliation{Email: lchotor33@yahoo.com}
\date{\today}

\begin{abstract}
In this work we study possibility of chaos formation in the
dynamics governed by paradigmatic model of Cavity Quantum
Electrodynamics, the so called James-Cammings model. In particular
we consider generalized JC model. It is shown that even in the
case of zero detuning dynamics is chaotic. Kinetic approach for
the problem under study has been applied.

\end{abstract}

\pacs{73.23.--b,78.67.--n,72.15.Lh,42.65.Re}
\maketitle


\section{Introduction}

\begin{figure}[t]
  \centering
  \includegraphics[width=6cm]{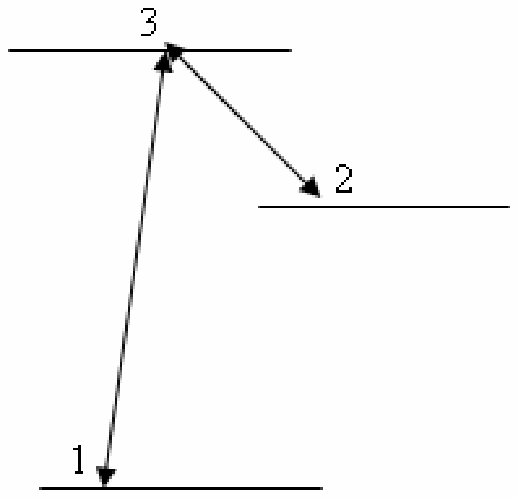}
  \caption{Schematic representation of atomic levels and available inter level transitions for
           generalized JC model.}\label{fig:1}
\end{figure}
Cavity quantum electrodynamics (CQED) is a rapidly developing
field of physics studying the interaction of atoms with photons in
the high-finesse cavities \cite{Aoki,Mabuchi,Hood,Raimond}.
Interest to such a systems basically is caused by two facts: One
of them is the possibility of more deep understanding of quantum
dynamics of open systems. Second argument is the possibility of
practical application in the field of quantum computing
\cite{Turchette}.

In particular CQED experiments implement a situation so simple
that their results are of great importance for better
understanding of fundamental postulates of quantum theory
\cite{Wineland}. They are thus appropriate for tests of basic
quantum properties: quantum superposition, complementarily or
entanglement. In the context of quantum information processing,
the atom and cavity are long-lived qubits, and their mutual
interaction provides a controllable entanglement mechanism an
essential requirement for quantum computing
\cite{Wineland,Monroe}.

In general dissipation processes must be taken into account when
discussing problems of   CQED. In particular there are two
dissipative channels for system: the atom may spontaneously emit
into modes other then preferred cavity mode, and photons may pass
through the cavity output coupling mirror.

But modern experiments in CQED have achieved strong atom-field
coupling for the strength of the coupling exceeds both decay
processes \cite{Ye,van Enk,Munstermann}. If so, then problem is
reduced to the driven Jaynes-Cummings (JC) Hamiltonian, which
models the interaction of a single mode of an optical cavity
having resonant frequency with a two level atom comprised of a
ground and exited states  \cite{Schleich}.

This model is basic model of interaction of radiation with matter
and describes the energy exchange between atom and quantized
radiation field in an ideal lossless cavity.

In further was offered JC model generalized for three level
optical atoms (see Fig.\ref{fig:1}). For more details see
\cite{Yoo,Bogoliubov}.

In most general case atom-radiation field interaction should
involve not only the internal atomic transitions and field states
but also the center-of-mass motion of the atom. With the recoil
effect taken into account, generalized JC Hamiltonian takes form
\be \hat{H}=\frac{\hat{p}^{2}}{2m}+\sum\limits_{\alpha =
1}^2\hbar\omega_{\alpha}\hat{b}_{\alpha}^{+}\hat{b}_{\alpha}+\sum\limits_{j=1}^3E_{j}\hat{R}_{jj}
+\hbar\sum\limits_
{\alpha=1}^2\hat{g}_{\alpha}(\hat{b}_{\alpha}\hat{R}_{3\alpha}+\hat{b}_{\alpha}^{+}\hat{R}_{\alpha3
})\cos(xk_{f_{\alpha}}) \ee

here $\hat{x}$ and  $\hat{p}$ are coordinate and pulse of the atom
$R_{3\alpha}$, $R_{\alpha3}$ are transition operators $\alpha=1,2$
, $\hat{R_{jj}}$ is the operator of the level population,
satisfying the condition $\sum\limits_{j=1}^{3}R_{jj}=1$. In the
basis of atomic states

$|a_{1}\rangle=\left(\begin{array}{l}
 0  \\
 0  \\
 1 \\
 \end{array} \right)$,
 $|a_{2}\rangle=\left(\begin{array}{l}
 0  \\
 1  \\
 0 \\
 \end{array} \right)$,
 $|a_{3}\rangle=\left(\begin{array}{l}
 1  \\
 0  \\
 0 \\
 \end{array} \right)$,
 $R_{jj}$~~~operators are the generators of the $SU(3)$  group
 $|\hat{R}_{11}\rangle=\left(\begin{array}{l}
 0~0~0  \\
 0~0~0  \\
 0~0~1 \\
 \end{array} \right)$,
 $|\hat{R}_{12}\rangle=\left(\begin{array}{l}
 0~0~0  \\
 0~0~0  \\
 0~1~0 \\
 \end{array} \right)$
 $,|\hat{R}_{13}\rangle=\left(\begin{array}{l}
 0~0~0  \\
 0~0~0  \\
 1~0~0 \\
 \end{array} \right)$,
 $|\hat{R}_{21}\rangle=\left(\begin{array}{l}
 0~0~0  \\
 0~0~1  \\
 0~0~0 \\
 \end{array} \right),$
 $|\hat{R}_{22}\rangle=\left(\begin{array}{l}
 0~0~0  \\
 0~1~0  \\
 0~0~0 \\
 \end{array} \right),$
 $|\hat{R}_{23}\rangle=\left(\begin{array}{l}
 0~0~0  \\
 1~0~0  \\
 0~0~0 \\
 \end{array} \right)$,
 $|\hat{R}_{31}\rangle=\left(\begin{array}{l}
 0~0~1  \\
 0~0~0  \\
 0~0~0 \\
 \end{array} \right)$,
 $|\hat{R}_{32}\rangle=\left(\begin{array}{l}
 0~1~0  \\
 0~0~0  \\
 0~0~0 \\
 \end{array} \right)$,

 $|\hat{R}_{33}\rangle=\left(\begin{array}{l}
 1~0~0  \\
 0~0~0  \\
 0~0~0 \\
 \end{array} \right)$.

Model given by the Hamiltonian (1) is nonlinear and non-integrable
and due to this may display chaotic behavior. Analogous problem
but for ordinary two level JC model was studied in \cite{Prants}
in detail. Result obtained in \cite{Prants} is that dynamics
essentially depends on detuning parameter between inter level
transition frequency and frequency of radiation field
$\delta=\omega_{\alpha}-E/\hbar$. Namely, in case of zero detuning
system displays regular behavior. Goal of this work is the study
of nonlinear dynamics governed by Hamiltonian (1). For better
understanding of the problem we shall discuss resonant case. We
shall try to find out difference in dynamics between generalized
and ordinary JC model studied in \cite{Prants}.

In resonant case, complete set of Heisenberg equations of motion
corresponding to  the Hamiltonian (1) looks like

\be
\begin{array}{l}
 \frac{dx(\tau)}{d\tau}=\alpha~p(\tau),  \\
 \frac{dp(\tau)}{d\tau}=u_{1}(\tau)\sin(x)+\Omega~ku_{2}(\tau)\sin(kx),  \\
 \frac{dR_{1}(\tau)}{d\tau}=\nu_{1}(\tau)\cos(x), \\
 \frac{dR_{2}(\tau)}{d\tau}=\Omega~\nu_{2}(\tau)\cos(kx),\\
 \frac{d\nu_{1}(\tau)}{d\tau}=2\cos(x)(M_{1}+1)(1-2R_{1}(\tau)-R_{2}(\tau))-\Omega \cos(kx)\cdot B(\tau),\\
 \frac{d\nu_{2}(\tau)}{d\tau}=2\Omega \cos(kx)(M_{2}+1)(1-R_{1}(\tau)-2R_{2}(\tau))-\cos(x)\cdot B(\tau),\\
 \frac{du_{1}(\tau)}{d\tau}=\Omega \cos(kx)\cdot C(\tau),\\
 \frac{du_{2}(\tau)}{d\tau}=-\cos(kx)\cdot C(\tau),\\
 \frac{dB(\tau)}{d\tau}= \cos(kx)(M_{1}+1)\nu_{2}(\tau)+\Omega \cos(kx)(M_{2}+1)\nu_{1}(\tau),\\
 \frac{dC(\tau)}{d\tau}=\cos(x)(M_{1}+1)u_{2}(\tau)-\Omega \cos(kx)(M_{2}+1)u_{1}(\tau)\\
 \frac{dN_{1}(\tau)}{d\tau}=\cos(x)\cdot~\nu_{1}(\tau),\\
 \frac{dN_{2}(\tau)}{d\tau}=\Omega \cos(kx)\cdot~\nu_{2}(\tau)\\
 \end{array} \ee

In (2) transform to the Hermit variables
$\hat{A}_{1}=\imath(\hat{b}_{1}\hat{R}_{31}-\hat{b}^{+}_{1}\hat{R}_{13})$,
$\hat{A}_{2}=\imath(\hat{b}_{2}\hat{R}_{32}-\hat{b}^{+}_{2}\hat{R}_{23})$,
$\hat{U_{1}}=(\hat{b}_{1}
\hat{R}_{31}+\hat{b}^{+}_{1}\hat{R}_{13})$,
$\hat{U_{2}}=(\hat{b}_{2}
\hat{R}_{32}+\hat{b}^{+}_{2}\hat{R}_{23})$,
$\hat{B}=\hat{b}_{1}\hat{b}^{+}_{2}\hat{R}_{21}+\hat{b}^{+}_{1}\hat{b}_{2}\hat{R}_{12}$,
$\hat{C}=\imath(\hat{b}_{1}\hat{b}^{+}_{2}\hat{R}_{21}-\hat{b}^{+}_{1}\hat{b}_{2}\hat{R}_{12})$,
$\hat{N}_{1}=\hat{b}^{+}_{1}\hat{b}_{1}$,$\hat{N}_{2}=\hat{b}^{+}_{2}\hat{b}_{2}$,

      and procedure of semi-classical averaging is done \cite{Prants}
 $x=k_{f_{1}}<\hat{x}>$,~~
 $p=\frac{<\hat{P}>}{\hbar
k_{f_{1}}}$,\\
~~$R_{1}=<R_{11}>$,~~$R_{2}=<R_{22}>$,~~$\nu_{1}=<\hat{A_{1}}>$,~~$\nu_{2}=<\hat{A_{2}}>$,
~~$u_{1}= \left<\hat{U}_{1}\right>$,~~$u_{2}=
\left<\hat{U}_{2}\right>$,~~$k=\frac{k_{f_{2}}}{k_{f_{1}}}$,
~~$B=\left<\hat{B}\right>$,~~$C=\left<\hat{C}\right>$, ~~$\tau =
g_{1}t$,~~$\alpha=\frac{k^{2}_{f_{1}}\hbar}{g_{1}m}$,~~$\Omega=\frac{g_{2}}{g_{1}}$
. In addition following notations for motion integrals are
introduced
$M_{1}=\hat{N}_{1}-\hat{R}_{11}=const$,~~$M_{2}=\hat{N}_{2}-\hat{R}_{22}=const$.

Below the set of equations (2) will be the object of our interest.
We shall consider two cases:

1)  Variable  $x$ is slow as against
$u_{1}(\tau),~~u_{2}(\tau)$.~~ In this case set of equations (2)
separates into two subsystems

\be
\begin{array}{l}
 \frac{du_{1}(\tau)}{d\tau}=\Omega \cos(kx)\cdot C(\tau)  \\
 \frac{du_{2}(\tau)}{d\tau}=-\cos(x)\cdot C(\tau),  \\
 \frac{dC(\tau)}{d\tau}=\cos(x)(M_{1}+1)u_{2}(\tau)-\Omega \cos(kx)(M_{2}+1)u_{1}(\tau), \\
 \frac{dx(\tau)}{d\tau}=\alpha p(\tau),\\
 \frac{dp(\tau)}{d\tau}=u_{1}(\tau)\sin(x)+\Omega ku_{2}(\tau)\sin(kx).\\
 \end{array} \ee
and
\be
\begin{array}{l}
 \frac{dR_{1}(\tau)}{d\tau}=\nu_{1}(\tau) \cos(x), \\
 \frac{dR_{2}(\tau)}{d\tau}=\Omega \nu_{2}(\tau) \cos(kx), \\
 \frac{d\nu_{1}(\tau)}{d\tau}=2\cos(x)(M_{1}+1)(1-2R_{1}(\tau)-R_{2}(\tau))-\Omega \cos(kx)B(\tau), \\
 \frac{d\nu_{2}(\tau)}{d\tau}=2\Omega \cos(kx)(M_{2}+1)(1-R_{1}(\tau)-2R_{2}(\tau))-\cos(kx)B(\tau),\\
 \frac{dB(\tau)}{d\tau}=\cos(x)(M_{1}+1)\nu_{2}(\tau)+\Omega cos(kx)(M_{2}+1)\nu_{1}(\tau),\\
 \frac{dN_{1}(\tau)}{d\tau}=\cos(x)\cdot \nu_{1}(\tau),\\
 \frac{dN_{2}(\tau)}{d\tau}=\Omega \cos(kx)\cdot \nu_{2}(\tau),\\
 \end{array} \ee

After solving of first three equations of (3), we have

\be
\begin{array}{l}
\ C(\tau)=A\sin(\Omega,\tau),\\
u_{1}(\tau)=u_{1}(0)-\frac{A\Omega}{\Omega_{1}}\cos(\Omega_{1}\tau),\\

u_{2}(\tau)=u_{2}(0)+\frac{A\Omega}{\Omega_{1}}\cos(\Omega_{1}\tau),\\

\end{array} \ee

where
$$\Omega^{2}_{1}=\cos^{2}(x)(M_{1}+1)+\Omega^{2}\cos^{2}(kx)(M_{2}+1).$$

  Then if,~~$k=\frac{k_{f_{1}}}{k_{f_{2}}}=1$  , motion of the atom inside of quantum cavity satisfies the equation

\be \frac{d^{2}x}{d\tau^{2}}
+|\alpha|(u_{1}(0)+u_{2}(0))\sin(x)=0,
\ee

with the possible solutions
  \be
  \frac{dx}{d\tau} =p=2\Theta \omega_{0}\left \{\begin{array}{ll}
   cn(\tau,\Theta), \ \ & \mbox{$\Theta\leq1$} \\
   dn(\tau,1/\Theta), \ \ & \mbox {$\Theta\geq1$}
  \end{array}\right.
,\ee
where
 $\Theta=\frac{1}{2}\left(1+H/\omega^{2}_{0}\right)$,~~
 $\omega^{2}_{0}=|\alpha|(u_{1}(0)+u_{2}(0))$,~~ $H=\frac{p^{2}(0)}{2}+\omega^{2}_{0}\cos(x_{0})$
 and $cn(...)$,~~$dn(...)$ are Jacobi elliptic functions \cite{Handbook}.
Two solutions (7) correspond to the two different phase
trajectories. One of those solutions  $dn(...)$  corresponds to
the closed phase trajectory. Other one corresponds to open phase
trajectory. In case of closed phase trajectory it means that after
the time interval , atom iterates initial state. For solving of
the subsystem (4) we can use the method given in
\cite{Bogoliubov}. As a result we get:

\be
\begin{array}{l}
\ R_{1}(\tau)=\mu(\cos(\lambda \tau)-1)+\beta \sin(\lambda
\tau)+\lambda^{2}_{1}[u(\cos(2\lambda \tau)-1)+\nu \sin(2\lambda
\tau )]+R_{1}(0),\\
\ R_{2}(\tau)=-\mu(\cos(\lambda \tau)-1)-\beta \sin(\lambda
\tau)+\lambda^{2}_{2}[u(\cos(2\lambda \tau)-1)+\nu \sin(2\lambda
\tau )]+R_{2}(0)\\
\end{array}
\ee

where $$\mu = \lambda^{-4}[
\lambda^{2}[\lambda^{2}_{2}R_{1}(0)-\lambda^{2}_{1}R_{2}(0)]
+(\lambda^{2}_{2}-\lambda^{2}_{1})K],~~\lambda=\sqrt{\lambda^{2}_{1}+\lambda^{2}_{2}},$$
$$\beta=\lambda^{-3}[\lambda^{2}_{2} \dot{R}_{1}(0)-\lambda^{2}_{1}
\dot{R}_{2}(0)], ~~~
u=\frac{1}{2}\lambda^{-4}[\lambda^{2}[2R_{1}(0)+2R_{2}(0)-1]+K],$$
$$\nu=\frac{1}{2}\lambda^{-3}[\dot{R}_{1}(0)+\dot{R}_{2}(0)],~~\lambda_{1}=\cos(x)\sqrt{M_{1}+1},
~~\lambda_{2}=\Omega\cos(kx)\sqrt{M_{2}+1},$$
and
$$N_{1}(\tau)=\mu (\cos(\lambda \tau)-1)+\beta \sin(\lambda \tau)-
\lambda^{2}_{1}[u(\cos(2\lambda \tau)-1)+\nu \sin(2\lambda
\tau)]+N_{1}(0),$$
$$N_{2}(\tau)= -\mu (\cos(\lambda \tau)-1)-\beta \sin(\lambda
\tau)+\lambda^{2}_{2}[u(\cos(2\lambda \tau)-1)+\nu \sin(2\lambda
\tau)]+N_{2}(0).$$

\begin{figure}[t]
  \centering
  \includegraphics[width=10cm]{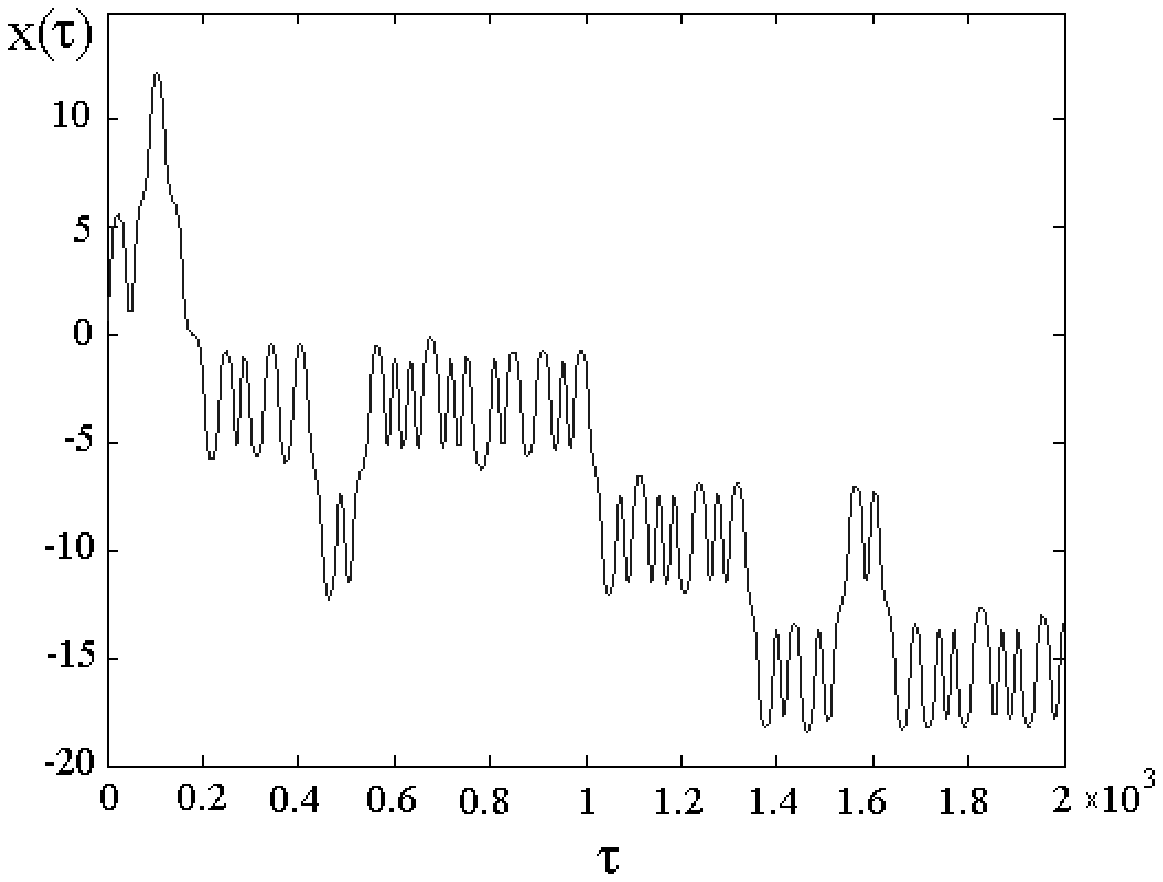}
  \caption{Plot of atom mass center coordinate dependence on time. As one can see from this Fig. motion
  is chaotic. As every other numeric result, plot is obtained for the values of
  parameters $\alpha=0.01$,~~$\Omega=0.5$,~~k=3,~~$M_{1}=1.6$,~~$M_{2}=1.7$}\label{fig:2}
\end{figure}

\begin{figure}[t]
  \centering
  \includegraphics[width=10cm]{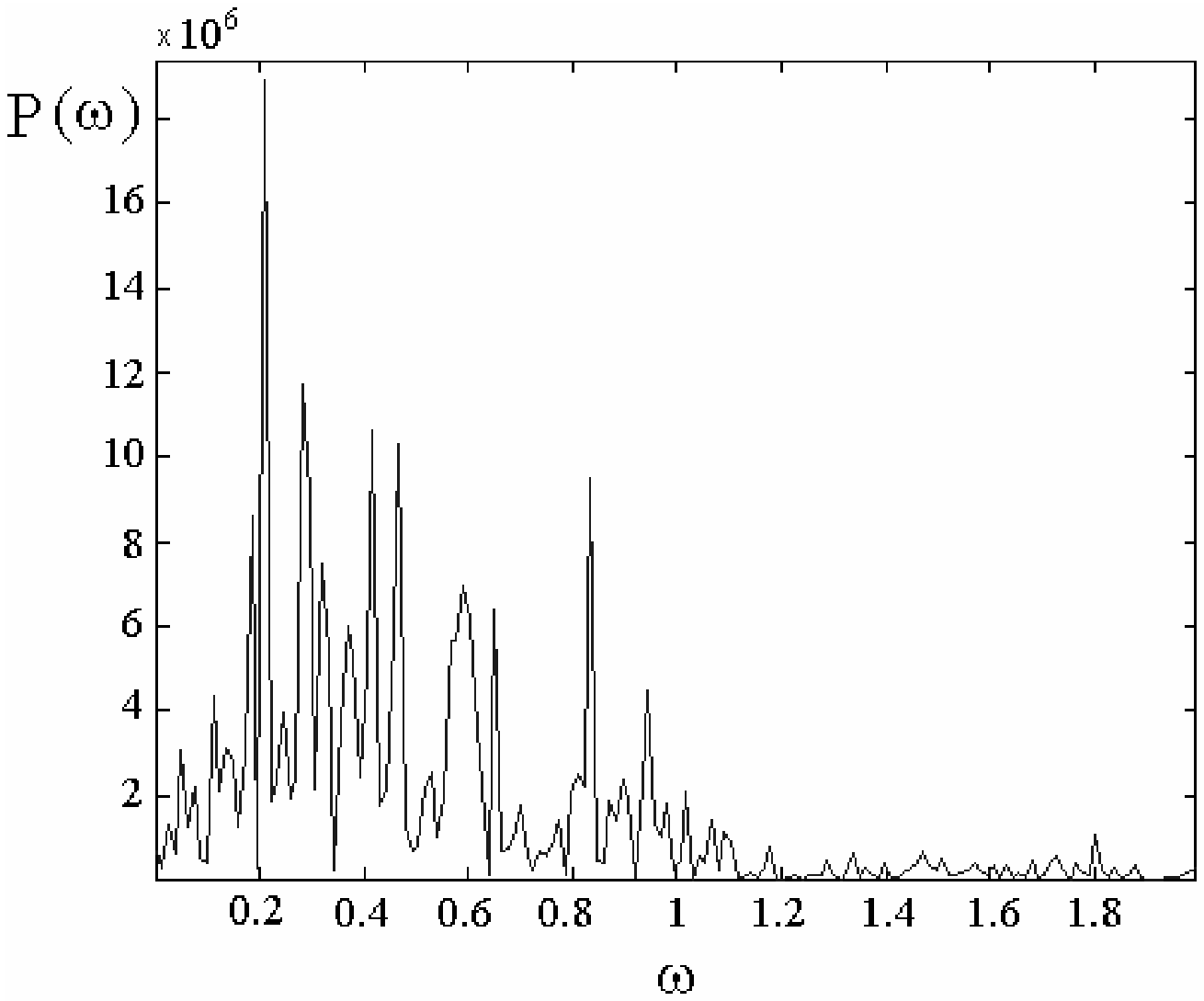}
  \caption{Fourier transform of the correlation function\\$ p(\omega)=G_{p}(\omega)=\int\limits_{0}^{\infty}d\tau
 \exp(\imath \omega \tau)G_{p}(\tau)$.~~Nonzero width of the correlation function is the sign of
 classical dynamical stochastisity} \label{fig:3}
\end{figure}

Really the quantities
$R_{1}(\tau)$,~~$R_{2}(\tau)$,~~$R_{3}(\tau)=1-R_{1}(\tau)-R_{2}(\tau)$
are observable on the experiment. But in general case, when
variable $x(\tau)$ is not adiabatic, we have to investigate (2) by
use of numerical methods. If $x(\tau)$ is non-adiabatic variable,
origination of chaos in the system is quite possible. In this case
dynamical consideration loses sense and transition from dynamic to
the statistical description is needed. As the result, statistical
conceptions, namely Kolmogorov entropy and fractional dimension
becomes important. Those concepts being statistical are strongly
related to the systems motional characteristics, namely to the
local instability of phase trajectories
\cite{Zaslavsky,Afraimovich,Lichtenberg}. Results of numeric
calculations, for the realistic values of parameters corresponding
to the high-finesse Fabry-Perot cavity and real atoms
\cite{Ye,Munstermann} are presented on
Fig.\ref{fig:2},~~Fig.\ref{fig:3}.

For determination of the width of Furrier transform of the
correlation function we used method of fast Furrier transform

$$G_{x}(\tau)=<x(t+\tau)x(t)>$$

$$G_{x}(\omega)=\int\limits_{0}^{\infty}d\tau G_{x}(\tau)\exp(\imath \omega \tau)
=\frac{\tau_{c}}{1+\omega^{2}\tau_{c}^{2}}$$

where
$<(...)>=\lim\limits_{T\rightarrow\infty}\frac{1}{T}\int\limits_{0}^{T}(...)dt$
means time average, $\tau_{c}$ is the correlation time. Result of
numeric calculations are presented on ~~Fig.\ref{fig:3}. Nonzero
width of the Fourier form is the sing of chaos.

One more sign for confirmation of chaos existence is the values of
maximal Lyapunov exponent. In this sense it is worth to compare
our result with the result obtained for ordinary JC model
\cite{Prants}. For numerical calculation of maximal Lyapunov
exponent we shall use G. Benettin's algorithm consisting in the
following procedures\cite{Lichtenberg}: Solving  two different
systems starting from the initial points
$x_{0}$,~~$x_{0}+\bar{x}_{0}^{0}$,~~$|\bar{x}_{0}^{0}|=\varepsilon,$
where $\varepsilon$ is small initial distance between phase
trajectories. Then after each step $x(t_{i})=x_{i}$,
$\bar{x}(t_{i})=x_{i}$ one has to do re-scaling
$\bar{x}_{i}^{0}=\varepsilon \bar{x}_{i}/|\bar{x}_{i}|$ use as
initial conditions $x_{i},x_{i}+\bar{x}_{i}^{0}$, and so on up to
the $x_{N},x_{N}+\bar{x}_{N}^{0}$ . Then Lyapunov exponent is
given via

\be
 \Lambda=\frac{1}{N}\sum\limits_{k=1}^{\N}\ln\left(\frac{\bar{x}_{k}}{\varepsilon}\right).
 \ee

Result of numeric calculations for the maximal Lyapunov exponent
is presented on Fig.4.
\begin{figure}[t]
  \centering
  \includegraphics[width=10cm]{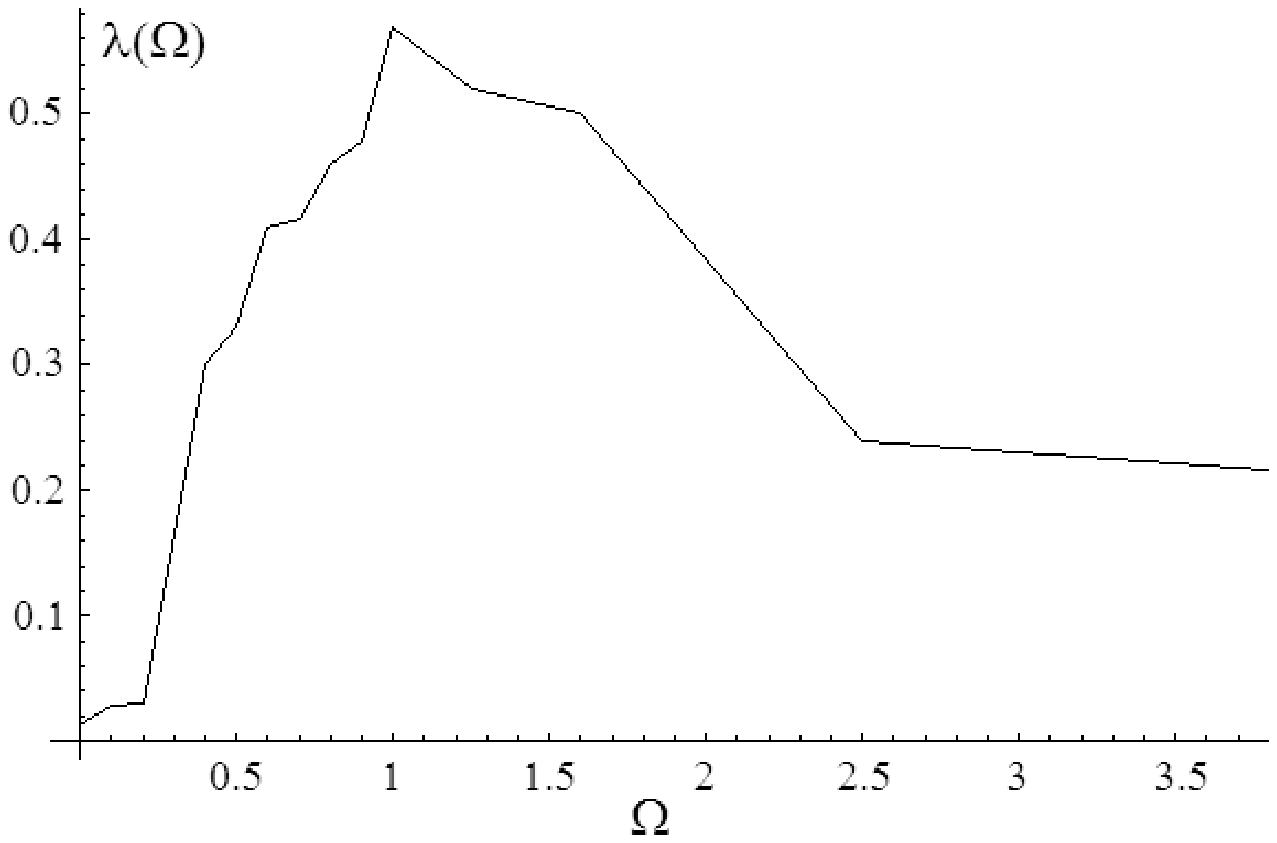}
  \caption{Maximal Lyapunov exponent as a function of ratio between constants of interaction
   between atom and radiation field $(\Omega=g_{2}/g_{1})$} \label{fig:4}
\end{figure}

It is evident from the Fig.\ref{fig:4}, that in case of ordinary
JC model corresponding to the values of parameter $\Omega=0$,
maximal Lyapunov exponent is equal to zero $\lambda=0$. This
result is in a good agreement with the results obtained in
\cite{Prants}.

 Other important physical quantity
characterizing classical dynamic stochasticity is the fractional
dimension of the systems phase space. For defining of fractional
dimension we will use algorithm of P.Grassberger, I.Procaccia
\cite{Grassberger,Grassberger1,Procaccia}.

\begin{figure}[t]
  \centering
  \includegraphics[width=10cm]{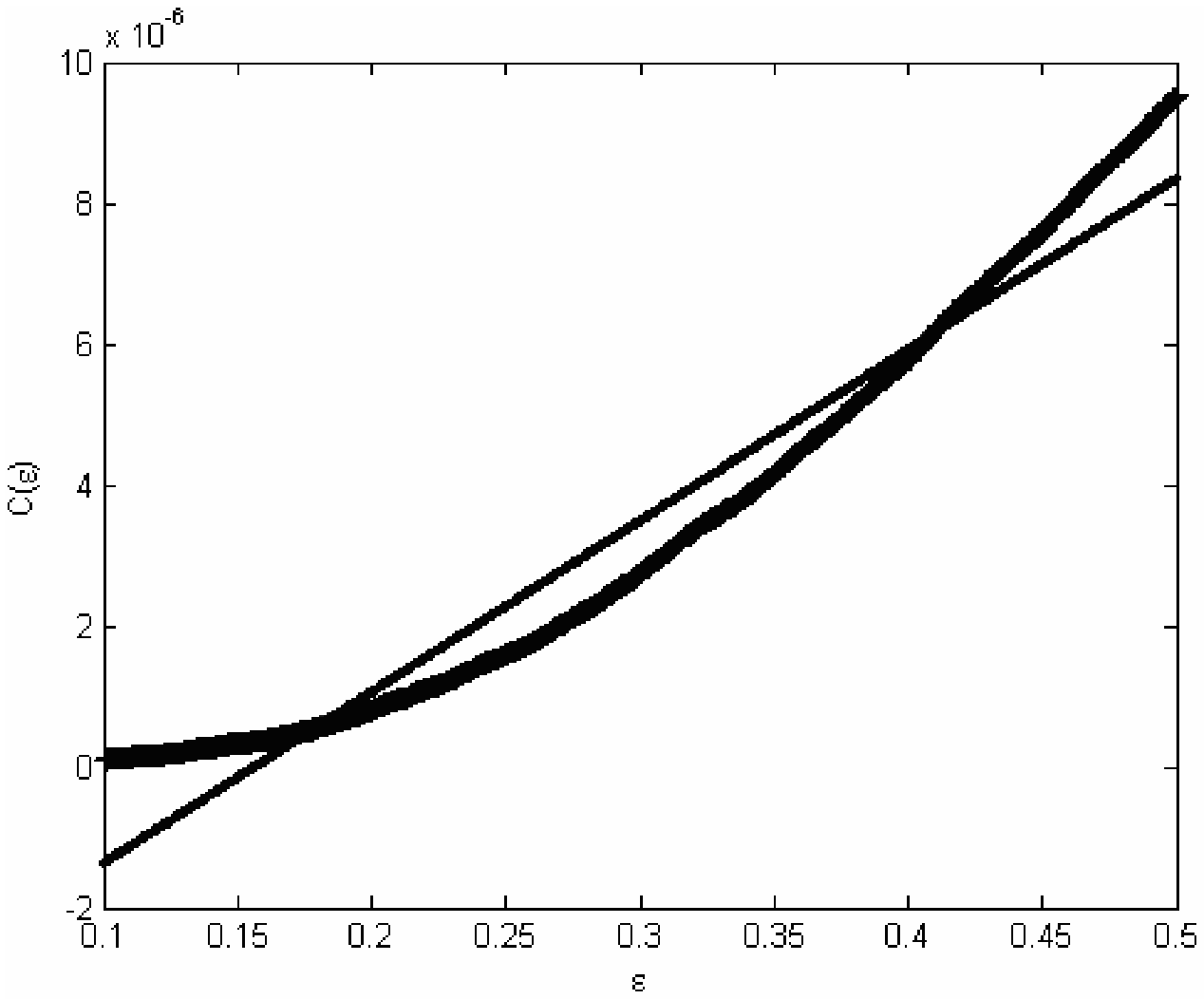}
  \caption{ Dependence of  $C(\varepsilon)$ on the values of $\varepsilon$ ,
  plotted by numerical integration of the set of equations (2) , for the following values of parameters
  $\alpha=0.01$, $\Omega=0.5$, $k=3$, $M_{1}=1.6$,$M_{2}=1.7$.
  A solid line corresponds to least-squares approximation of the
results of data processing.
  According to this plot,
  fractal dimension of the system
  $D=\frac{\ln C(\varepsilon_{2})-\ln C(\varepsilon_{1})}{\ln(\varepsilon_{2})-\ln(\varepsilon_{1})}$
  is equal to
  $D\approx2.44$} \label{fig:5}
\end{figure}

Let we have set of state vectors
($\chi_{i},i=1,2,\cdots,N$)corresponding to the successive steps
of numeric integration. In our case $\chi_{i}$ is the complete set
of variables (2) with the values corresponding to the moments of
time $t=t_{i}$. Then we can use numeric data for estimation of
following expression
$$C(\varepsilon)=\lim_{N\to\infty}\frac{1}{N(N-1)}\sum_{i, j=1}^{N}\theta(\varepsilon-\vert\chi_{i}-\chi_{j}\vert)$$
where $ \theta(x)=\left \{ \begin {array}{ll}
                          0   & \mbox{$x<0$} \\
                          1   & \mbox{$x\geq0$}
                          \end{array}
                          \right. $
 is the step function. According to the P.Grassberger, I.Procaccia \cite{Grassberger,Grassberger1,Procaccia}
 fractional dimension may be defined as
 $$D=\lim_{\varepsilon\to0}\frac{C(\varepsilon)}{\log(\varepsilon)}$$
 Expected dependence of $C(\varepsilon)$ is $\varepsilon^{D}.$
 So, plot must be a line with angular coefficient $D$.
 Result of numeric calculations are presented on ~~Fig.\ref{fig:5}.

 \section{Quantum mechanical Consideration. Mixed State Formation.}
 On the basis of numeric results of previous section, we can conclude that  dynamics is
  chaotic. In case of quantum consideration we neglect kinetic energy of atomic
  motion as compared to the level transition frequencies, and random nature of motion we shall try
  to take into account by considering $x(t)$ as time dependent random Wiener process.
  As a result in interaction representation we get:
  \be
  i\frac{d\vert\psi(t)\rangle}{dt}=\widehat{V}\vert\psi(t)\rangle
  \ee

where interaction operator is \be
\widehat{V}=\hbar\sum_{\alpha=1}^{2}g_{\alpha}(\widehat{b}_{\alpha}\widehat{R}_{3\alpha}+
\widehat{b}_{\alpha}^{+}\widehat{R}_{\alpha 3})\cdot\cos
k_{f_{\alpha}}\widehat{x} \ee Let present initial wave function as
a direct product of atomic and field states
$$\vert\psi(0)\rangle=\vert\psi_{atom}\rangle\otimes\vert\psi_{field}\rangle $$
where
$$ \vert\psi_{field}\rangle=\sum_{n_{1}=0}^{\infty}W_{n_{1}}\vert n_{1}\rangle+
\sum_{n_{2}=0}^{\infty}W_{n_{2}}\vert n_{2}\rangle,
\vert\psi_{atom}\rangle=C_{a_{1}}\vert a_{1}\rangle+C_{a_{2}}\vert
a_{2}\rangle+C_{a_{3}}\vert a_{3}\rangle$$
$$\vert a_{1}\rangle=\left( \begin{array}{l}
    0 \\ 0 \\1 \end{array} \right),
    \vert a_{2}\rangle=\left( \begin{array}{l}
    0 \\ 1 \\0 \end{array} \right),
    \vert a_{3}\rangle=\left( \begin{array}{l}
    1 \\ 0 \\0 \end{array} \right).$$
    As far as operator (11) can mix only following states $\vert a_{3} \   n_{1}\rangle,$
    $\vert a_{1}  \ n_{1}+1\rangle,$ $\vert a_{3}\  n_{2}\rangle,$ $\vert a_{2} \  n_{2}+1\rangle,$
    solution of (10) we will search in the form
    \be
    \vert\psi(t)\rangle=\sum_{a_{j}n_{i}}C_{a_{j}n_{i}}(t)\vert
    a_{j}n_{i}\rangle.
    \ee
    After substitution (12) in (10) and solving equations for
    $C_{a_{j}n_{i}}(t)$ we get
    \be
    \begin{array}{l}
    \ C_{a_{3}n_{1}}(t)=\frac{C_{1}}{2}e^{-i\lambda_{1}\int\limits_{0}^{t}{\cos
    k_{f_{1}}x(t')dt'}}+\frac{C_{2}}{2}e^{i\lambda_{1}\int\limits_{0}^{t}{\cos
    k_{f_{1}}x(t')dt'}},\\
    \ C_{a_{1}n_{1}+1}(t)=\frac{C_{1}}{2}e^{-i\lambda_{1}\int\limits_{0}^{t}{\cos
    k_{f_{1}}x(t')dt'}}-\frac{C_{2}}{2}e^{i\lambda_{1}\int\limits_{0}^{t}{\cos
    k_{f_{1}}x(t')dt'}}.
    \end{array}
    \ee
    where $\lambda_{1}=g_{1}\sqrt{n_{1}+1}$
    $$C_{1}=C_{a_{3}n_{1}}(0)+C_{a_{1}n_{1}+1}(0); \ \ \ C_{2}=C_{a_{3}n_{1}}(0)-C_{a_{1}n_{1}+1}(0);$$
\be
    \begin{array}{l}
    \ C_{a_{3}n_{2}}(t)=\frac{C_{3}}{2}e^{-i\lambda_{2}\int\limits_{0}^{t}{\cos
    k_{f_{2}}x(t')dt'}}+\frac{C_{4}}{2}e^{i\lambda_{2}\int\limits_{0}^{t}{\cos
    k_{f_{2}}x(t')dt'}},\\
    \ C_{a_{2}n_{2}+1}(t)=\frac{C_{3}}{2}e^{-i\lambda_{2}\int\limits_{0}^{t}{\cos
    k_{f_{2}}x(t')dt'}}-\frac{C_{4}}{2}e^{i\lambda_{2}\int\limits_{0}^{t}{\cos
    k_{f_{2}}x(t')dt'}}.
    \end{array}
    \ee
 Here $\lambda_{2}=g_{2}\sqrt{n_{2}+1}$
$$C_{3}=C_{a_{3}n_{2}}(0)+C_{a_{2}n_{2}+1}(0); C_{4}=C_{a_{3}n_{2}}(0)-C_{a_{2}n_{2}+1}(0);$$
Density matrix of the system: field + atom is direct product of
two density matrixes
 \be
 \rho_{ijkl}=\rho_{atom}\otimes\rho_{field}=\overline{C_{a_{i}n_{j}}\cdot C_{a_{k}n_{l}}^{\ast}}
\ee where $\overline{(\ldots)}$ means averaging.

Sign of mixed state formation is zeroing of non-diagonal matrix
elements of density matrix (15), \cite{Landau,Feynman}. This
process is non-reversible because, along with zeroing of
non-diagonal matrix elements, loss of information about wave
function's phase factor takes place. So, to prove irreversibility
of dynamics, it is sufficient to show zeroing of non-diagonal
matrix elements.

Substituting (13), (14) into (15), it is easy to see that
non-diagonal matrix elements contain terms like
 \be
C_{a_{i}n_{j}}(0)C_{a_{k}n_{l}}^{\ast}(0)\exp
\left[i\int\limits_{0}^{t}{\omega_{1,2}^{\pm}(t')dt'}\right] \ee
where $\omega_{1,2}^{\pm}(t')=g_{1}\cos (k_{f_{1}}x(t'))\pm g_{2}
\cos(k_{f_{2}}x(t'))$. Due to the random nature of $x(t'),$ below
we will consider $\omega_{1,2}^{\pm}$ as a random process and for
simplicity omit indices. It is clear that exponent in (16) may be
considered as a functional of the random function $\omega(t)$ \be
Q[\omega]=\exp\left[i\int\limits_{0}^{t}\omega(t')dt'\right]. \ee
So, to get non-diagonal matrix elements of the density matrix
(15), we need to do statistical averaging with respect to the all
possible realizations of random function $\omega(t)$ \be
\rho_{ijkl, \atop i\neq k,j\neq l}\approx
C_{a_{i}n_{j}}(0)C_{a_{k}n_{l}}^{\ast}(0)\left\langle
Q[\omega[t]]\right\rangle \ee where
$\left\langle(\ldots)\right\rangle$ means statistical average.
Average values of the functional (18) may be calculated by doing
following continual integral \be
Q[\omega]=\exp\left[i\int\limits_{0}^{t}{\omega(t')dt'}\right]=
\lim\limits_{N\rightarrow\infty \atop \Delta
t_{k}\rightarrow0}\int{d\omega_{N}\ldots
d\omega_{1}\exp\left[i\sum_{k=1}^{n}\omega_{k}\Delta
t_{k}\right]P_{N}[\omega]}\ee where $\Delta
t_{k}=t^{(k)}-t^{(k-1)},$ $t^{(0)}=0,$  $t^{(N)}=1,$
$P_{N}[\omega]$ is the multi-dimensional normal distribution
function given by \be
P_{N}[\omega]=(2\pi)^{-N}\int{d\lambda_{1}\ldots
d\lambda_{N}\exp\left[-i\sum_{k}{\lambda_{k}\omega_{k}}\right]\exp\left[-\frac{1}{2}\sum_{k,k^{'}}{C_{kk^{'}}\lambda_{k}\lambda_{k^{'}}}\right]}\ee
where $\lambda_{k}$ are the distribution parameters and
$C_{kk^{'}}$ is the covariation matrix \cite{Feller}

After substituting (20) in (19) and doing integration we obtain
\be
 \begin{array}{l}
 \ \int{d\omega_{1}\ldots d\omega_{N}\exp[i\sum\limits_{k=1}^{N}\omega_{k}\Delta
 t_{k}]P_{N}[\omega]}=\\
 =\int {d\lambda_{1}\ldots
 d\lambda_{N}\exp\left[-\frac{1}{2}\sum\limits_{k,k^{'}}C_{k,k^{'}}\lambda_{k}\lambda_{k^{'}}\right]}\times
 \prod\limits_{k=1}^{N}\left\{\frac{1}{2\pi}\int{\exp[i\omega_{k}(\Delta
 t_{k}-\lambda_{k})]d\omega_{k}}\right\}=\\
 =\int{d\lambda_{1}\ldots d\lambda_{N}
 \delta(\lambda_{1}-\Delta t_{1})\delta(\lambda_{2}-\Delta t_{2})\ldots\delta(\lambda_{N}-\Delta t_{N})
 \times\exp\left[-\frac{1}{2}\sum\limits_{k,k^{'}}\Delta t_{k}\Delta t_{k^{'}}\right]}
 \end{array}
    \ee
    So, for average value of the functional we get
    \be
    \left\langle Q[\omega]\right\rangle=\lim\limits_{N\rightarrow\infty \atop\Delta t_{k}\rightarrow
    0}\exp\left[-\frac{1}{2}\sum\limits_{k,k^{'}}c(t^{(k)},t^{(k^{'})})\Delta t_{k}\Delta t_{k^{'}}\right]=
    \exp\left[-\frac{1}{2}\int\limits_{0}^{t}dt^{'}\int\limits_{0}^{t^{''}}{c(t^{'},t^{''})}\right] \ee
    For normal random process $c(t^{'},t^{''})=c(t^{'}-t^{''}).$ Then making transform
    to the new variables $t^{'}-t^{''}=\tau,$ $t^{'}+t^{''}=\xi$  and doing integration in (22) over variable
    $\xi$ we get
    \be
    \left\langle
    Q[\omega]\right\rangle=\exp\left[-\frac{1}{2}t\int\limits_{-t}^{t}{d\tau}c(\tau)\right].
    \ee
    After assuming that correlation function
    $c(\tau)=\left<\omega(t+\tau)\omega(t)\right>$ has the Gaussian form $c(\tau)=e^{-\alpha_{0}\tau^{2}},$
    from (23) we get
    \be
    \left\langle
    Q[\omega]\right\rangle\approx\exp\left[-\sqrt{\frac{\pi}{\alpha_{0}}}\frac{t}{2}\cdot
    Erf[t\sqrt{\alpha_{0}}]\right]
    \ee
    where $Erf(\ldots)$ is the error function \cite{Feller}.

    So, for non-diagonal matrix elements we have
    \be \rho_{ijkl \atop i\neq k,j\neq l}\approx
     C_{a_{i}n_{j}}(0)C_{a_{k}n_{l}}^{\ast}(0)
    \exp\left[-\sqrt{\frac{\pi}{\alpha_{0}}}\frac{t}{2}\cdot Erf[t\sqrt{\alpha_{0}}]\right].\ee

From (25) it is clear that after the lapse of time
$t>\sqrt{\alpha_{0}/\pi}$ zeroing of non-diagonal matrix elements
and formation of mixed state happens.

 \section{Kinetic Description}
 In previous section we investigated mechanism of mixed state formation.
 After mixed state is formed, quantum-mechanical consideration loses meaning and
 there is a need to use kinetic description. Kinetic equation for chaotic quantum-mechanical
 system first time was obtained in \cite{Zaslavsky}. But this study was done in the
 semi-classical domain. Namely, in \cite{Zaslavsky} zeroing of non-diagonal part of density matrix was proved
 by use of semi-classical approximation. Our purpose is to do the same in exceptionally
 quantum domain, without application to the semi-classical methods. As the zeroing of
 non-diagonal part is already shown, problem is partly solved.

 It's clear that system atom$+$ field is adiabatically isolated.
 So entropy product is zero. But by considering field as a thermostat,
 we can calculate entropy growth for atomic subsystem.If $\Delta S>0,$
 then process in non-reversible.\\
 As we have already mentioned at $t=0,$ atom is not connected with light and state vector
 $\vert\psi(t=0)\rangle$ is the direct product of two states
 \be
 \vert\psi(0)\rangle=\vert\psi_{atom}\rangle\otimes\vert\psi_{field}\rangle
 \ee
 where
 \be
 \vert\psi_{field}\rangle=\sum\limits_{n_{1}=0}^{\infty}{W_{n_{1}}\vert n_{1}\rangle}+
 \sum\limits_{n_{2}=0}^{\infty}{W_{n_{2}}\vert n_{2}\rangle},\ \ \ \vert\psi_{atom}\rangle=
 C_{a_{1}}\vert a_{1}\rangle+C_{a_{2}}\vert a_{2}\rangle+C_{a_{3}}\vert a_{3}\rangle
\ee
 Let us note that interaction (11) mixes only states
 $$\vert a_{3} \ n_{1}\rangle,\ \ \ \vert a_{1} \ n_{1}+1\rangle,\ \ \ \vert
 a_{3} \ n_{2}\rangle,\ \ \
 \vert a_{2} \ n_{2}+1\rangle.$$
 Then comparing (26) with (12) one can see that initial conditions in (13), (14) are of form
 \be
 \begin{array}{l}
 \ C_{a_{3}n_{1}}(0)=W_{n_{1}}C_{a_{3}},\\
 \ C_{a_{1}n_{1}+1}(0)=W_{n_{1}+1}C_{a_{1}},\\
 \ C_{a_{3}n_{2}}(0)=W_{n_{2}}C_{a_{3}},\\
 \ C_{a_{2}n_{2}+1}(0)=W_{n_{2}+1}C_{a_{2}}.
 \end{array}
 \ee
Let us remind that inversion of atomic populations $I$ is a value
of interest, since it can be measured on the experiment. Inversion
is defined as difference between level populations
\be
 \begin{array}{l}
 \ I_{a_{3}a_{2}}(t)=W(t,\vert a_{3}\rangle)-W(t,\vert
 a_{2}\rangle),\\
 \ I_{a_{3}a_{1}}(t)=W(t,\vert a_{3}\rangle)-W(t,\vert
 a_{1}\rangle),
\end{array}
 \ee
 where
 \be
 W(t,\vert a\rangle)=\sum\limits_{n=0}^{\infty}\vert
 C_{an}(t)\vert^{2}.
 \ee
 After substituting (13), (14) in (30), and using initial conditions (28) we get
 \be
 \begin{array}{l}
 \ W(t,\vert a_{3}\rangle)=\frac{1}{4}\sum\limits_{n_{2}=0}^{\infty}\left\vert(W_{n_{2}}C_{a_{3}}+
 W_{n_{2}+1}C_{a_{2}})Q_{n_{2}}^{\ast}[\omega_{2}]+(W_{n_{2}}C_{a_{3}}-W_{n_{2}+1}C_{a_{2}})Q_{n_{2}}[\omega_{2}]\right\vert^{2}+
 \\
\ + \frac{1}{4}\sum\limits_{n_{1}=0}^{\infty}\left\vert
(W_{n_{1}}C_{a_{3}}+W_{n_{1}+1}C_{a_{1}})Q_{n_{1}}^{\ast}[\omega_{1}]+(W_{n_{1}}C_{a_{3}}-W_{n_{1}+1}C_{a_{1}})Q_{n_{1}}[\omega_{1}]\right\vert^{2},\\
\ W(t,\vert
a_{2}\rangle)=\frac{1}{4}\sum\limits_{n_{2}=0}^{\infty}\left\vert(W_{n_{2}}C_{a_{3}}+
 W_{n_{2}+1}C_{a_{2}})Q_{n_{2}}^{\ast}[\omega_{2}]-(W_{n_{2}}C_{a_{3}}-W_{n_{2}+1}C_{a_{2}})Q_{n_{2}}[\omega_{2}]\right\vert^{2},\\
\ W(t,\vert
a_{1}\rangle)=\frac{1}{4}\sum\limits_{n_{1}=0}^{\infty}\left\vert
(W_{n_{1}}C_{a_{3}}+W_{n_{1}+1}C_{a_{1}})Q_{n_{1}}^{\ast}[\omega_{1}]-(W_{n_{1}}C_{a_{3}}-W_{n_{1}+1}C_{a_{1}})Q_{n_{1}}[\omega_{1}]\right\vert^{2}.
\end{array}
 \ee
 To get we also have introduced following notation for unitary functional
  \be
  Q_{n_{1,2}}[\omega_{1,2}]=\exp\left[-ig_{1,2}\sqrt{n_{1,2}+1}\int\limits_{0}^{t}
  {\cos(k_{f_{1,2}}x(t^{'}))dt^{'}}\right],
  \ee
  \be
  Q_{n_{1,2}}^{\ast}[\omega_{1,2}]=Q_{n_{1,2}}^{-1}[\omega_{1,2}],\ \ \
  Q_{n_{1,2}}[\omega_{1,2}]Q_{n_{1,2}}^{-1}[\omega_{1,2}]=1.
  \ee
  By using of (31) it is easy to define inversion values for any moment of time.
   But before doing this let us mention one interesting fact.  In ordinary JC model
   which for zero detuning is not characterized by chaos, inversion is characterized by periodical
   revivals in time. See for example \cite{Averbukh}.
   So absence of periodical revivals maybe useful for experimental observation of quantum chaos.
    After using (32), (33) from (31) we get
   \be
  \left \langle W(t,\vert
   a_{2}\rangle)\right\rangle=\sum\limits_{n_{2}=0}^{\infty} \big[\frac{1}{2}(W_{n_{2}}^{2}C_{a_{3}}^{2}+W_{n_{2}+1}^{2}C_{a_{2}}^{2})-
   \frac{1}{4}(W_{n_{2}}^{2}C_{a_{3}}^{2}-W_{n_{2}+1}^{2}C_{a_{2}}^{2})(\langle
   Q_{n_{2}}^{2}[\omega_{2}]\rangle+\langle
   Q_{n_{2}}^{-2}[\omega_{2}]\rangle)\big]
   \ee
 \be
  \left \langle W(t,\vert
   a_{1}\rangle)\right\rangle=\sum\limits_{n_{1}=0}^{\infty}\big[\frac{1}{2}(W_{n_{1}}^{2}C_{a_{3}}^{2}+W_{n_{1}+1}^{2}C_{a_{1}}^{2})-
   \frac{1}{4}(W_{n_{1}}^{2}C_{a_{3}}^{2}-W_{n_{1}+1}^{2}C_{a_{2}}^{2})(\langle Q_{n_{1}}^{2}[\omega_{1}]\rangle+\langle Q_{n_{1}}^{-2}[\omega_{1}]\rangle)\big]
   \ee
   where $\left<\ldots\right>$ again means statistical average.
     Then taking into account(24)

from (34), (35) we get \be
\begin{array}{l}
\ \left\langle W(t,\vert
a_{2}\rangle)\right\rangle=\sum\limits_{n_{2}=0}^{\infty}[\frac{1}{2}(W_{n_{2}}^{2}C_{a_{3}}^{2}+
W_{n_{2}+1}^{2}C_{a_{2}}^{2})],\\
\ \left\langle W(t,\vert
a_{1}\rangle)\right\rangle=\sum\limits_{n_{1}=0}^{\infty}[\frac{1}{2}(W_{n_{1}}^{2}C_{a_{3}}^{2}+
W_{n_{1}+1}^{2}C_{a_{1}}^{2})],\\
\ \left\langle W(t,\vert a_{3}\rangle)\right\rangle=\left\langle
W(t,\vert a_{1}\rangle)\right\rangle+\left\langle W(t,\vert
a_{2}\rangle)\right\rangle.
\end{array}
\ee

At last let us remind that $W_{n_{1,2}}^{2}$ describes field
states and obeys to Poisson distribution \cite{Loudon}
 \be
W_{n_{1,2}}^{2}=\frac{\bar{n}_{1,2}^{n_{1,2}}\exp(-\bar{n}_{1,2})}{n_{1,2}!},\ee
and $C_{a_{1}},$ $C_{a_{2}},$ $C_{a_{3}}$ are probabilities of
level occupations.

   Let us assume that, at the initial moment of time, atom is in the lowest state
 \be
\rho_{a_{1}}(t=0)=\left\vert
C_{a_{1}n_{1}+1(t=0)}\right\vert^{2}=1,\ \ \
\rho_{a_{2}}(t=0)=\rho_{a_{3}}(t=0)=0. \ee Then initial values of
entropy is zero \be
S(t=0)=\sum\limits_{a_{i}=1}^{3}\rho_{a_{i}}(t=0)\ln(\rho_{a_{i}}(t=0))=0,\ee
so system is in pure quantum-mechanical state.

After the lapse of time $t=t_{0}$ more than  time of inter level
transition $t_{0}\sim 1/g_{\alpha}$, system may  perform multiple
transitions between levels. That is why probability to find system
in other states will be nonzero: \be C_{a_{1}}\neq
0,~~C_{a_{2}}\neq 0,~~C_{a_{3}}\neq 0,~~~~~t>t_{0}.\ee
 Despite of this fact to talk about probability of population of different states
is early yet. The point is that in time interval:
 \be
t_{0}<t<\sqrt{\frac{\alpha_{0}}{\pi}} \ee interferentional terms
are nonzero. Therefore the state of the system will be pure one.
But unlike of the initial state (38),which is simple state, the
state of the system in time interval (41) is superposition one.

Superposition state is pure quantum mechanical state and only
after zeroing of interferentional terms in (18) superposition
state passes to mixed one. Such a transition occurs in times: \be
t>\sqrt{\frac{\alpha_{0}}{\pi}}\ee But in time interval (41) while
the system is in pure superposition state, from the symmetry point
of view, it is clear that the coefficient values(40) have to
satisfy the following relation:
 \be
C_{a_{1}}\left(t_{0}<t<\sqrt{\frac{\alpha_{0}}{\pi}}\right)\sim
C_{a_{2}}\left(t_{0}<t<\sqrt{\frac{\alpha_{0}}{\pi}}\right)\sim
C_{a_{3}}\left(t_{0}<t<\sqrt{\frac{\alpha_{0}}{\pi}}\right)\sim C.
\ee Before proceed to the more exact determination of the
quantities (36), we note one important fact.
    Values of the parameters (28) $C_{a_{1}},$  $C_{a_{2}},$  $C_{a_{3}},$
 taken at the initial moment of time $t=0,$ define level populations.
 At the moment of time $t=0$ atom and cavity field are not connected with each other (26).
 Therefore the following relation should be hold:
 \be
 C^{2}_{a_{1}}+ C^{2}_{a_{2}}+ C^{2}_{a_{3}}=1.
 \ee
After the laps of time interaction (11) will mix atom and field's
states and due to the fact that (41) does not correspond to the
normalization condition of complete wave function, it will not be
valid henceforth. Considering of level populations again is
possible only after the time interval (42). But in this case level
populations are defined by the quantities (36), (28) and not by
the coefficients (69). fficients (69).
      For determination of level's populations we shall use (36),(43)
 and take into account that fields states $W_{n_{1,2}}$
 obeys Poisson distribution (37). As a result we obtain

Then taking into account (36)as a result we get \be
\rho_{a_{1}}=\frac{1}{4}, \ \ \ \rho_{a_{2}}=\frac{1}{4},\ \ \
\rho_{a_{3}}=\frac{2}{4},\ee Result is very interesting. Most
populated is high exited level $\rho_{a_{3}}.$ This means that we
have stochastic absorption of field energy. For entropy growth we
have \be \Delta S=\frac{1}{2}(\ln4+\ln2)>0\ee So, process is
non-reversible.
\section{Conclusions}
Let sum up and analyze the results obtained in conclusion.

The aim of this work was to study generalized JC model being
subject to resonator field. Interest to such a systems is caused
by the fact that they are the most perspective to be used in
quantum computer. The question that came up is the following: by
how much will be state of the system controllable and dynamics
reversible? We have considered the most general case, when
interaction of the system with field depends on coordinate of the
system inside resonator.

Contrary to generally accepted opinion, it has turned out that the
absence of detuning, between resonator field and frequency of the
inter-level transitions, does not guarantee reversibility of the
system's state. During evolution in time the system executes
irreversible transition from pure quantum-mechanical state to
mixed one. At the same time, the time needed for formation of
mixed state $t>\sqrt{\alpha_{0}/\pi}$ is determined completely by
the autocorrelation function of random variable $\omega(t).$
Randomsity of this variable in it's turn is connected to the
chaotic motion of atoms inside of cavity.


\begin{thebibliography}{26}
\bibitem{Aoki} T. Aoki et al. Nature \textbf{443}, 671-674, (2006)
%
\bibitem{Mabuchi}H. Mabuchi, and A. Doherty, Science \textbf{298}, 1372, (2002)
%
\bibitem{Hood} C. J.Hood et al. Science \textbf{287}, 1447 (2000)
%
\bibitem{Raimond}J. Raimond, M. Brune, and S. Haroche, Rev. Mod. Phys. \textbf{73}, 565,(2001)
%
\bibitem{Turchette} Q. A. Turchette et al. Phys. Rev. Lett. \textbf{75}, 4710, (1995)
%
\bibitem{Wineland}D. J. Wineland et al. J. Res. Natl. Inst. Stand. Technol. \textbf{103}, 259,(1998)
%
\bibitem{Monroe} C. Monroe et al. Science \textbf{272}, 111 (1996)
%
\bibitem{Ye} J. Ye, D. W.Vemooy, and H. J. Kimble, Phys.Rev. Lett. \textbf{83}, 4987, (1999)
%
\bibitem{van Enk}S.J. van Enk, J. Mckeever, H. J. Kimble, and J. Ye. Phys.Rev. A. \textbf{64},013407,(2001)
%
\bibitem{Munstermann}P. Munstermann, T. Fischer, P. Maunz, P. W. H. Pinkse,and G. Rempe Phys. Rev. Lett.
\textbf{82}, 3791, (1999)
%
\bibitem{Schleich}P. Schleich \emph{Quantum Optics in Phase Space}, Wiley-VCH, Berlin,(2001)
%
\bibitem{Yoo}H. Yoo, J. Eberly  Phys. Rep. \textbf{v.118}, p. 240, (1985)
%
\bibitem{Bogoliubov}N.N. Bogoliubov, Jr., FamLe Kien, A.S. Shumovsky  Phys.Lett A. v.\textbf{101}, p.201, (1984)
%
\bibitem{Prants}S. Prants, N.Edelman, G.Zaslavsky,Phys.Rev. E\textbf{66}, 046222 (2002).
%
\bibitem{Handbook}\emph{Handbook of Mathematical Functions with Formulas, Graphs,
      and Mathematical Tables} National Bureau of Standards, Applied
      Mathematical Series, \textbf{55}, U.S. Government Printing Office,
      (Washington D.C., 1964)
%
\bibitem{Zaslavsky} G. M. Zaslavsky \emph{Physics of Chaos in Hamiltonian Systems},
     Imperial College Press, London, 1998.
%
\bibitem{Afraimovich} V. Afraimovich and S.-B. Hsu, \emph{Lectures on Chaotic Dynamical Systems},
American Mathematical Society, International Press, 2000.
%
\bibitem{Lichtenberg} A. J. Lichtenberg and M.A. Liberman, \emph{Regular and Chaotic Dynamics}
Springer-Verlang,Berlin, 1992
%
\bibitem{Grassberger} P.Grasberger, Phys.Lett.A,v.97,p.227,(1983).
%
\bibitem{Grassberger1} P.Grasberger, Phys.Lett.A,v.97,p.224,(1983).
%
\bibitem{Procaccia} P.Grasberger,I.Procaccia,Physica D,v.9,p.189,(1983).
%
\bibitem{Landau} L.D.Landau and E.M.Lifshitz, Quantum Mechanics, Non-relativistic Theory (Pergamon,Oxford,1977).
%
\bibitem{Feynman} R.P. Feynman, Statistical Mechanics (W.A.Benjamin,Inc. Massachusets, 1972)
%
\bibitem{Feller} W.Feller, An introduction to probability Theory
and Its Applications, John Wiley @ \textmd{\emph{sons}}, Inc. New
York, London, Sidney v.1 (1958), v.2 (1966).
%
\bibitem{Averbukh} Sh. Averbukh, Phys. Rev. A,v.46, p.2205,
(1992)
\bibitem{Loudon} R. Loudon The Quantum Theory of Light, Clarendon
Press, Oxford (1973)
%
\end{thebibliography}
\end{document}